\documentclass[aps,pra,twocolumn,showpacs,superscriptaddress,groupedaddress]{revtex4-1}  

\newcommand{\ket}[1]{\left\vert#1\right\rangle}

\newcommand{\Rb}{$^{87}Rb$}

\usepackage{physics}
\usepackage{graphicx}
\usepackage{epstopdf}
\usepackage{setspace}
\usepackage{bbold}
\usepackage{comment}
\usepackage{color}
\usepackage{soul}
\usepackage[table]{xcolor}
\usepackage{hyperref}
\usepackage{float}

\begin{document}

\author{Ido Almog, Jonathan Coslovsky, Gil Loewenthal, Arnaud Courvoisier and Nir Davidson}

\address{$^1$Department of Physics of Complex Systems, Weizmann Institute of Science, Rehovot 76100, Israel}
%

\title[]{High phase space density loading of a falling magnetic trap}

\begin{abstract}

Loading an ultra-cold ensemble into a static magnetic trap involves unavoidable loss of phase space density when the gravitational energy dominates the kinetic energy of the ensemble.
In such a case the gravitational energy is transformed into heat, making a subsequent evaporation process slower and less efficient.
We apply a high phase space loading scheme on a sub-doppler cooled ensemble of Rubidium atoms, with a gravitational energy much higher than its temperature of $1~\rm{\mu K}$. Using the regular configuration of a quadrupole magnetic trap, but driving unequal currents through the coils to allow the trap center to fall, we dissipate most of the gravitational energy and obtain a 20-fold improvement in the phase space density as compared to optimal loading into a static magnetic trap. Applying this scheme, we start an efficient and fast evaporation process as a result of the sub-second thermalization rate of the magnetically trapped ensemble.
\end{abstract}


\maketitle

\emph{Introduction - }Quantum degenerate gases~\cite{Anderson1995,pethick_smith_2008,Truscott2001} have proven to be essential tools in the exploration of fields such as condensed-matter physics~\cite{Bloch2005}, quantum simulation~\cite{Bloch2012}, study of phase transitions~\cite{Greiner2002} or topological properties of matter~\cite{Goldman2016}. Reaching quantum degeneracy in dilute gases requires numerous cooling stages~\cite{Phillips1997}, and their joint integration in a single experimental apparatus can become cumbersome.
A prominent example is the combination of Raman sideband cooling (RSBC)~\cite{Kerman2000} and conventional evaporation in a magnetic trap~\cite{Migdall1985,Ketterle1999}.
RSBC is simple to implement and provides large ($>10^8$) number of cold ($\sim1~\rm{\mu K}$) atoms at high phase space densities (PSD) $>10^{-4}$. However, as a conventional loading of RSBC cooled atoms into a magnetic trap leads to a drastic loss of PSD, this technique is barely used in cold-atom experiments. This issue stems from the gravitational energy $mg_0h$ of the ensemble (where $m$ is the mass of the atoms, $g_0\simeq9.8~\text{m/s}^{2}$ and $h\simeq1~\rm{mm}$ is the vertical extent of the cloud) which is typically $\sim 100$ times higher than its kinetic energy. In order to maintain a high PSD one needs to dissipate the high gravitational energy of the ensemble during the loading process. This is also applicable to Raman cooling \cite{Kasevich1992,Davidson1994} and velocity selective coherent population trapping techniques~\cite{Aspect1988,Aspect1989} as both schemes yield clouds whose gravitational potential energies are typically larger than their kinetic energies.

In this work we theoretically and experimentally show that it is possible to eliminate most of this excess gravitational energy by loading the atoms into a free-falling magnetic trap that is then adiabatically compressed and decelerated until it reaches a stop. The trap's fall and compression are controlled electronically with two magnetic coils in a quadrupole configuration driven with unbalanced current. Using this scheme we end up with a 20-fold increase in the PSD of the ensemble as compared to optimal loading into a static magnetic trap. Combining RSBC and efficient loading we are able to reach a PSD of $\sim 10^{-4}$, similar to more complex experimental setups \cite{streed2006,Linn2009}. Decelerating magnetic traps have been recently used to slow down cold but fast molecular beams without loss of PSD~\cite{Edvardas2011}.

\emph{Principle - } For the sake of explaining how to optimize such a falling magnetic trap, we start by neglecting gravity and consider a static trap that is suddenly turned on \cite{Ketterle1999}. If the density and velocity distributions of the ensemble have a Gaussian profile with root mean square extents $z_{\text{rms}}$ and $v_{\text{rms}}$ respectively, its loading into an isotropic harmonic trap with the oscillation frequency $\omega_{\text{osc}}= v_{\text{rms}}/z_{\text{rms}}$ preserves PSD. If the trap potential is linear as in the case of a quadrupole trap, the optimal trapping gradient to preserve PSD is of order of the kinetic energy of the ensemble divided by its size. Numerically it was found~\cite{Ketterle1999} to be $mb_{\text{opt}}\simeq3.7k_BT/z_{\text{rms}}$, where $m b_{\text{opt}}$ is the gradient of the quadrupole magnetic field along the tight axis $z$, while in the $x$ and $y$ directions, the gradient is smaller by a factor of two.
For our RSBC cooled ensemble with $T=1~\rm{\mu K}$ and $z_{\text{rms}}=600~\rm{\mu m}$ in all axes, the optimal gradient is $0.6~\text{m/s}^{2}$. It corresponds to $\sim 1/20$ of the free falling acceleration $g_0$ and therefore, the optimal gravity-free trap cannot support the ensemble against gravity and we are forced to use a much tighter trap, leading to significant heating of the cloud.


By allowing the trap center to fall and accelerate at $g_0$, the gravitational force is compensated by the fictitious d'Alembert force in the non-inertial frame of reference. Hence, a free-falling trap suddenly turned on with an initial optimal (weak) potential, then compressed and decelerated adiabatically until it stops will cause no deleterious heating. For a practical optimization of the trap parameters, gravity compensations and adiabaticity will only be approximated. Indeed, one has to take into account a finite fall constraint due to the limited size of any experimental apparatus, $13$~mm in our case.

\emph{Optimal motion profiles - } We derive  the equation of motion for a given finite fall. The symmetry axis of our quadrupole configuration is aligned along the gravitation field direction $-z$. The trapping potential near the quadrupole center reads
\begin{equation}
\label{eq_potential}
V(x,y,z)=\frac{1}{2} m b_z(t) \sqrt{x^2+y^2+4\left[z-z_c(t)\right]^2} \ \ ,
\end{equation}
where $b_z(t)$ and $z_c(t)$ are time dependent quantities describing the trap gradient and the trap center's position, respectively.

In order to ensure adiabatic compression of the cloud, the logarithmic change in the trap gradient should fulfill $d\log(b_z)(t)/dt=\tau^{-1}(t)$, where $\tau(t)=z_{\text{rms}}/v_{\text{rms}}$ is a typical adiabatic timescale. From the equipartition theorem, one gets that $b_zz_{\text{rms}}\propto k_BT\propto v_{\text{rms}}^2$. Hence, if the PSD is successfully kept constant ($v_{\text{rms}}z_{\text{rms}}=\text{const}$),  we easily get that $\tau(t)\sim b_z^{-2/3}(t)$. This yields the adiabatic condition
\begin{equation}
\label{eq_eq_diff}
\frac{d\log\pqty{b_z}}{dt} (t)=\gamma\tau_0^{-1}\pqty{\frac{b_z(t)}{b_0}}^{2/3},
\end{equation}where, $m b_0$ should be of the order of $m b_{\text{opt}}$ and $\tau_0\simeq z_{\text{rms}}(t=0)/v_{\text{rms}}(t=0)$ is the initial adiabatic timescale. The arbitrary numerical factor $\gamma$ is kept as a free parameter for simulations.
The solution to this differential equation reads
\begin{equation} 
\label{eq_slopetraj}
b_z(t)=b_0 \left(1-\frac{2\gamma t}{3\tau_0}\right)^{-3/2} \ \ .
\end{equation}The function, $b_z(t)$, diverges for $t\rightarrow 3\tau_0/2\gamma$, so it has to be clamped down before reaching any physical limit.

Next, we derive the trajectory of the trap center $z_c(t)$ using $b_z(t)$ as derived above.
The relation between the two quantities is best understood in the accelerating frame of the trap center. In this reference frame, gravity and d'Alembert's force result in an asymmetric trapping potential. The trap gradients above $b_z^+(t)$ and below $b_z^-(t)$ the center of the trap read 
\begin{equation}
b_z^\pm(t)=b_z(t)\pm\left(g_0+\frac{d^2z_c}{dt^2}(t)\right).
\end{equation}

\begin{figure}
\centering
\centerline{\includegraphics[trim={0.3cm 1.2cm 0.3cm 1.7cm},clip,width=9.5cm]{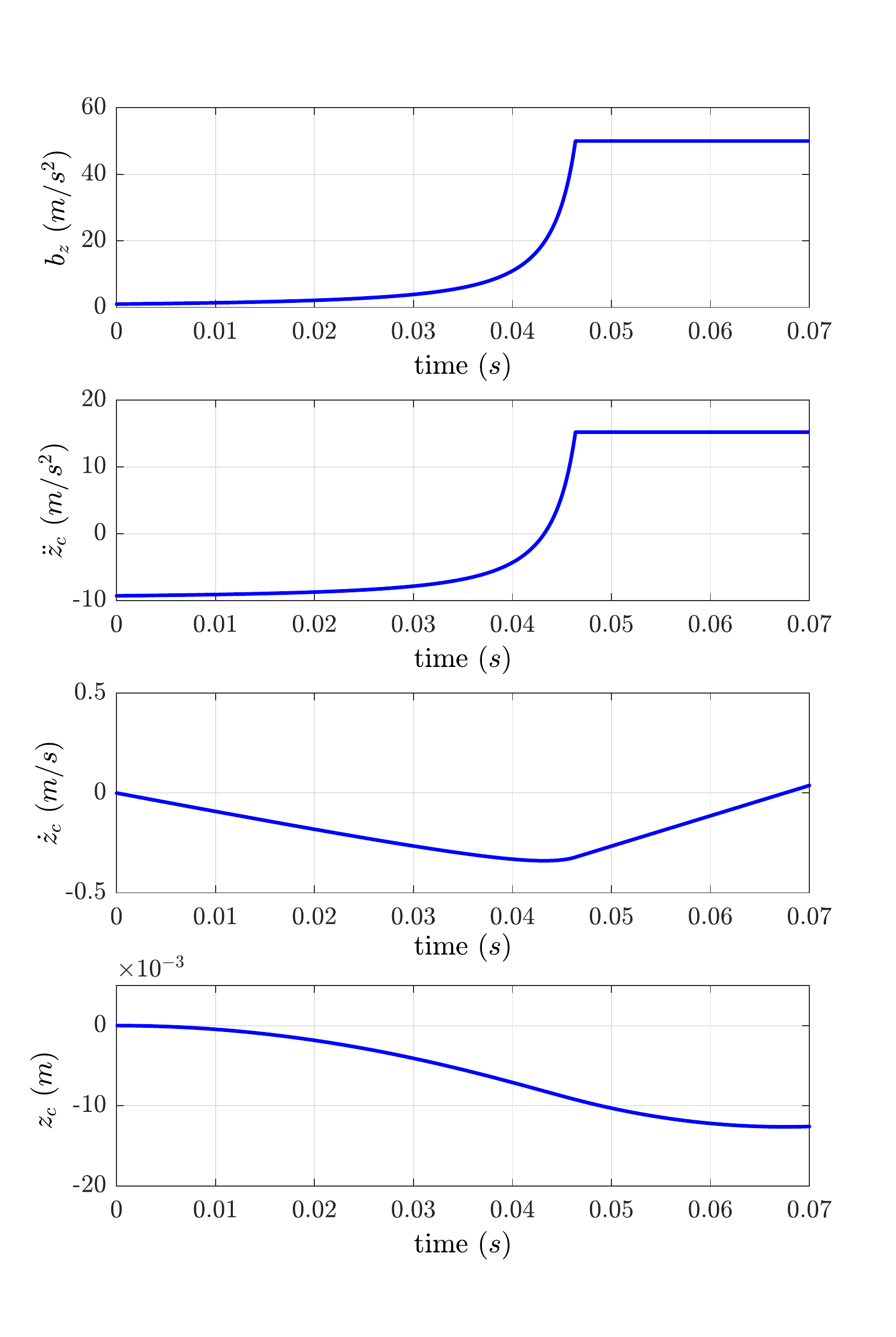}}
\caption{Optimal falling magnetic trap trajectories for conditions similar to that of the experiment where the ensemble is allowed to fall $13\:$mm before hitting the bottom of the cell. The trap parameters are $\tau_0=50~\rm msec$, $b_0=0.98~\rm m.s^{-2}$ and $\varepsilon = 1$. Here we chose $\gamma=3/2$. The trajectory acceleration is clamped at a maximum gradient of $b_{z,max}=50~\rm m.s^{-2}$, continued with a constant deceleration until the cloud reaches a stop, near the cell's bottom.}
\label{fig_traj}
\end{figure}

The asymmetry of the trap can be characterized by the parameter $\varepsilon~=\left(b_z^+(t)-b_z^-(t)\right)/b_z(t)$, which yields the following equation of motion for the trap center in the laboratory's rest frame 
\begin{equation} \label{eq_postraj}
\frac{d^2z_c}{dt^2}(t)=\frac{\varepsilon}{2} b_z(t)-g_0.
\end{equation}
For $\varepsilon>2$, in our case of an upward acceleration, the two gradients have opposite signs and there is no minimum to the trapping potential. The smaller $\varepsilon$ is the better the PSD is preserved but the total trap fall consequently increases. We find numerically that $\varepsilon = 1$ already yields satisfying results. The trajectories, obtained using equations \ref{eq_slopetraj} and \ref{eq_postraj} with $\varepsilon = 1$ and optimal parameters for $b_0$ and $\tau_0$ are shown in figure ~\ref{fig_traj}. The trap gradient is clipped to $50~\rm m.s^{-2}$ corresponding to the limiting current in our quadropole magnetic coils, thus also clipping the trap acceleration in order to maintain $\varepsilon = 1$. The trajectory used to stop the fall and bring the ensemble to rest in the trap's neutral center is easily kept adiabatic as the \textit{a priori} compression of the trap causes the adiabatic timescale to be short and easy to fulfill.

We run a Monte-Carlo simulation to solve Newton's equations for the motion of individual atoms in the trap, starting with the thermodynamic parameters of our sub-doppler cooled ensemble. By comparing the distribution in position and momentum space before and after the trap is set to a stop, we optimize the parameters while constraining the trap to a fall of $13\:$mm.
Starting with an initial guess for the parameters ($b_0=b_{\text{opt}}$, $\tau_0=z_{\text{rms}(0)}/v_{\text{rms}(0)}$, $\varepsilon = 1$, $\gamma=3/2$), we optimize the PSD with respect to all parameters. To get a point of comparison we do the same for a sudden turn on of a static magnetic trap.

The main results of the simulation are summarized in table \ref{tab_psd} and optimization curves of the initial magnetic field gradient are displayed in figure 2. For a static trap in the presence of gravity, the optimal loading maintains  $1\%$ of the initial PSD. By allowing the trap to fall $15\rm mm$ we maintain $27\%$ of the initial PSD. Without gravity (equivalent to an infinite fall with $\varepsilon \rightarrow 0$ and $\tau_0 \rightarrow \infty$) we maintain $\sim 73\%$ of the initial PSD, as in~\cite{Ketterle1999}.\begin{figure}
\centering
\centerline{\includegraphics[trim={0cm 0cm 0cm 0.89cm},clip,width=10cm]{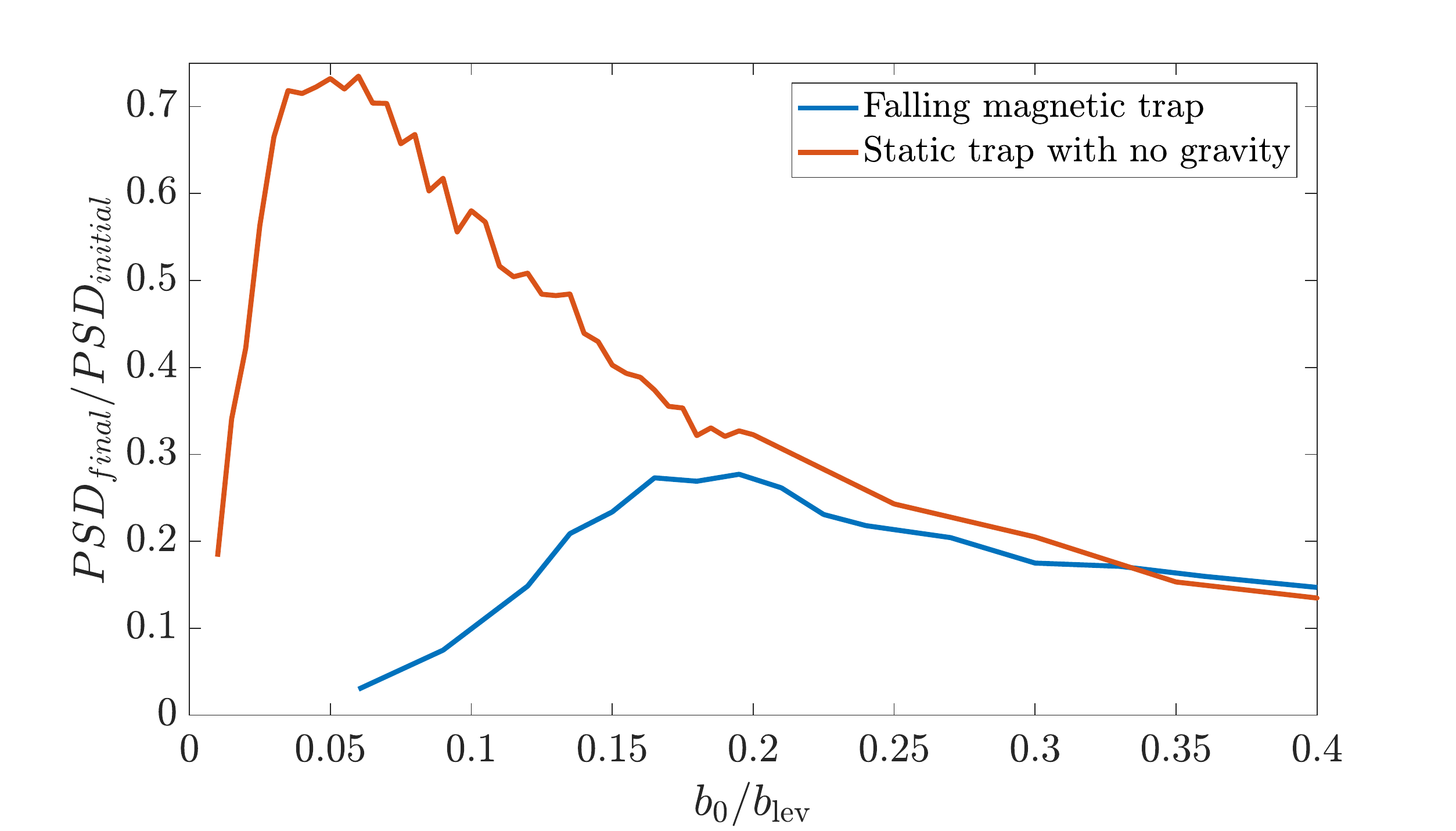}}
\caption{An example of Monte Carlo simulation with 2500 atoms showing the ratio between the final and initial PSDs as a function of $b_0/b_{lev}$, where $b_{lev}=15.25~\rm{G.cm^{-1}}$ is the levitation gradient. Are shown the cases of a static magnetic trap without gravity and that of a falling magnetic trap in the presence of gravity. For both simulations the cloud had a temperature of $1\mu K$ and a waist of $600\mu m.$ For the falling magnetic trap, we chose $\epsilon = 1.6$, $\gamma=3/2$ and a fall that was restricted to $15~\rm{mm}$.}
\label{}
\end{figure}

\begin{table}
\centering
\begin{tabular}{|l|c|c|c|c|}
  \hline
  PSD ($\times 10^{-6}$)& initial & no fall & $13~\rm mm$ fall & infinite fall \\
  \hline
  Simulation & $400$ & $4$ & $100$ & $300$ \\
  \hline
  Experiment & $400$ & $3$ & $53$ &  \\
  \hline
\end{tabular}
\caption{The PSD in simulation and experiment for different traps. The falling distance is the main factor to obtain a large PSD in the trap.}
\label{tab_psd}
\end{table}

\emph{Experiment - }Our single vacuum chamber experimental setup is described elsewhere \cite{Sagi2012} and the optical cooling stages are described here in short. Constant current flows through dispensers which evaporate the atoms into the chamber.
Rubidium $87$ atoms are loaded during 3 seconds into a standard magneto-optical trap (MOT), followed by $\sim 10\rm ms$ of Sisyphus cooling~\cite{Dalibard1989} using the MOT beams. Finally, we perform $\sim 10\rm ms$ of RSBC~\cite{Kerman2000} in a 3-dimensional optical lattice detuned by $+13\:$GHz from the  $5^2S_{1/2}$ to $5^2P_{3/2}$ transition. The RSBC optical pumping beam mainly $\sigma^+$ circularly polarized with respect to an applied $100~\rm mG$ magnetic field is detuned by $+10~\rm MHz$ from the transition between the $\ket{5^2S_{1/2},\: F=1}$ and $\ket{5^2 P_{3/2},\: F'=0}$ states.
At the end of this cooling stage the atoms end up in the $\ket{F=1,\: m=+1}$ Zeeman state and we transfer them to the magnetic trappable state, $\ket{F=2,\: m=2}$ by means of a constant micro-wave radiation, while ramping down the bias magnetic field.
We end up with $N=2\times10^8$ atoms in a spherically symmetric cloud of size $z_{\text{rms}}=600~\rm \mu m$ at $T=1~\rm \mu K$.
A PSD of $4\times 10^{-4}$ is calculated using $PSD=N\left(2\pi z_{\text{rms}}^2 \right)^{-3/2}\left(h/\sqrt{2\pi k_B T m}\right)^3$. This figure is $\sim100$ times higher than the typical $10^{-6}$ obtained after polarization gradient cooling in \Rb\ setups~\cite{Ketterle1999}.

\begin{figure}
\centering
\centerline{\includegraphics[trim={0.3cm 1.2cm 1.5cm 0.5cm},clip,width=9.5cm]{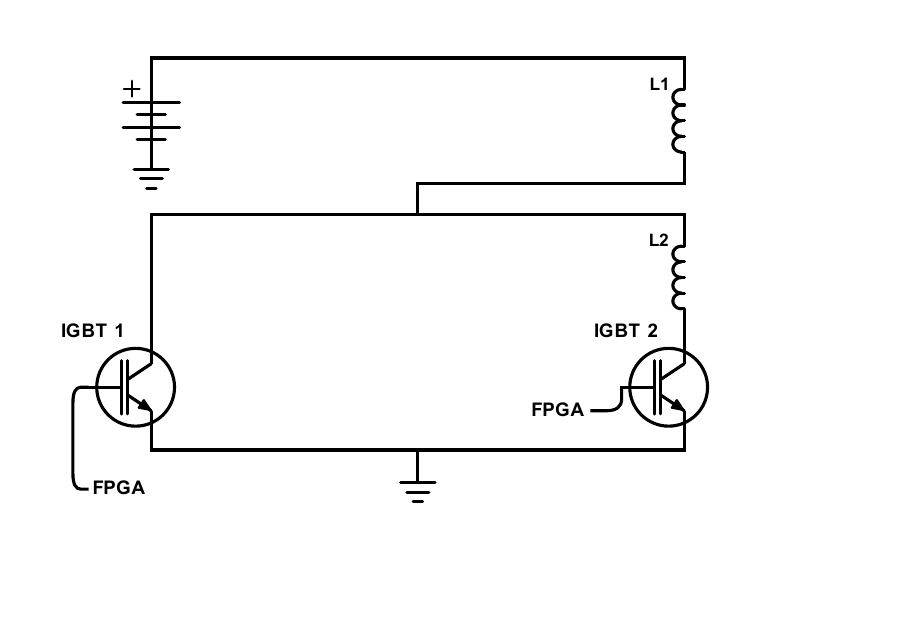}}
\caption{Simplified schematic of the electric circuit used to control the current flowing through the MOT coils $L_1$ and $L_2$. We use two IGBTs (IGBT 1 and IGBT 2) in order to control the ratio $I_{\text{L}_1}/I_{\text{L}_2}$ of the currents in the top and bottom coils, respectively}.
\label{fig:IGBT}
\end{figure}

The implementation of the falling magnetic trap requires analog control of the currents in the MOT coils while they are in an anti-Helmoltz configuration. As illustrated in figure \ref{fig:IGBT}, We use a pair of Powerex CM400HA-24H insulated-gate bipolar transistors (IGBT) whose gate-emitter voltages are controlled via analog outputs of an FPGA. The IGBTs are wired such that the current in the upper coil (L1) is the sum of the currents flowing through IGBTs 1 and 2, $I_{\text{L}_1}=I_{\text{IGBT1}}+I_{\text{IGBT2}}$. The current in the second coil is given by $I_{\text{L}_2}=I_{\text{IGBT2}}$. By using the FPGA analog output, one is thus able to easily control the ratio $I_{\text{L}_1}/I_{\text{L}_2}$.

The falling magnetic trap method is then applied according to the trajectory derived above. Conveniently, $b_z$ and $z_c$ are approximately linear functions of the currents through the coils.
While the field gradient depends on the average current, the position of the trap's minimum is a function of the difference between the two currents. At the end of the fall, near the cell floor, the ensemble is adiabatically brought back to its original position within $\sim 100~\rm ms$.
We then ramp up slowly the quadrupole current to produce a trap with a gradient of $b_z=87~\rm m.s^{-2}$, from which the ensemble is suddenly released after $0.5~\rm s$ of thermalization.
We use florescence imaging after a variable time of flight to measure the cloud's temperature and, using a fixed trap gradient we determine its PSD.
We optimize the trap's trajectory parameters, including a non-zero  initial trap velocity, in order to maximize the final PSD.
For the optimal falling trap parameters, the temperature along the $z$ axis (gravitation) is measured to be $80~\rm \mu K$, still higher than the temperature $58~\rm \mu K$ along the other axes,\begin{figure}[h!]
  \centering
  \includegraphics[trim={0 0 0 -1.7cm}, width=0.45\textwidth]{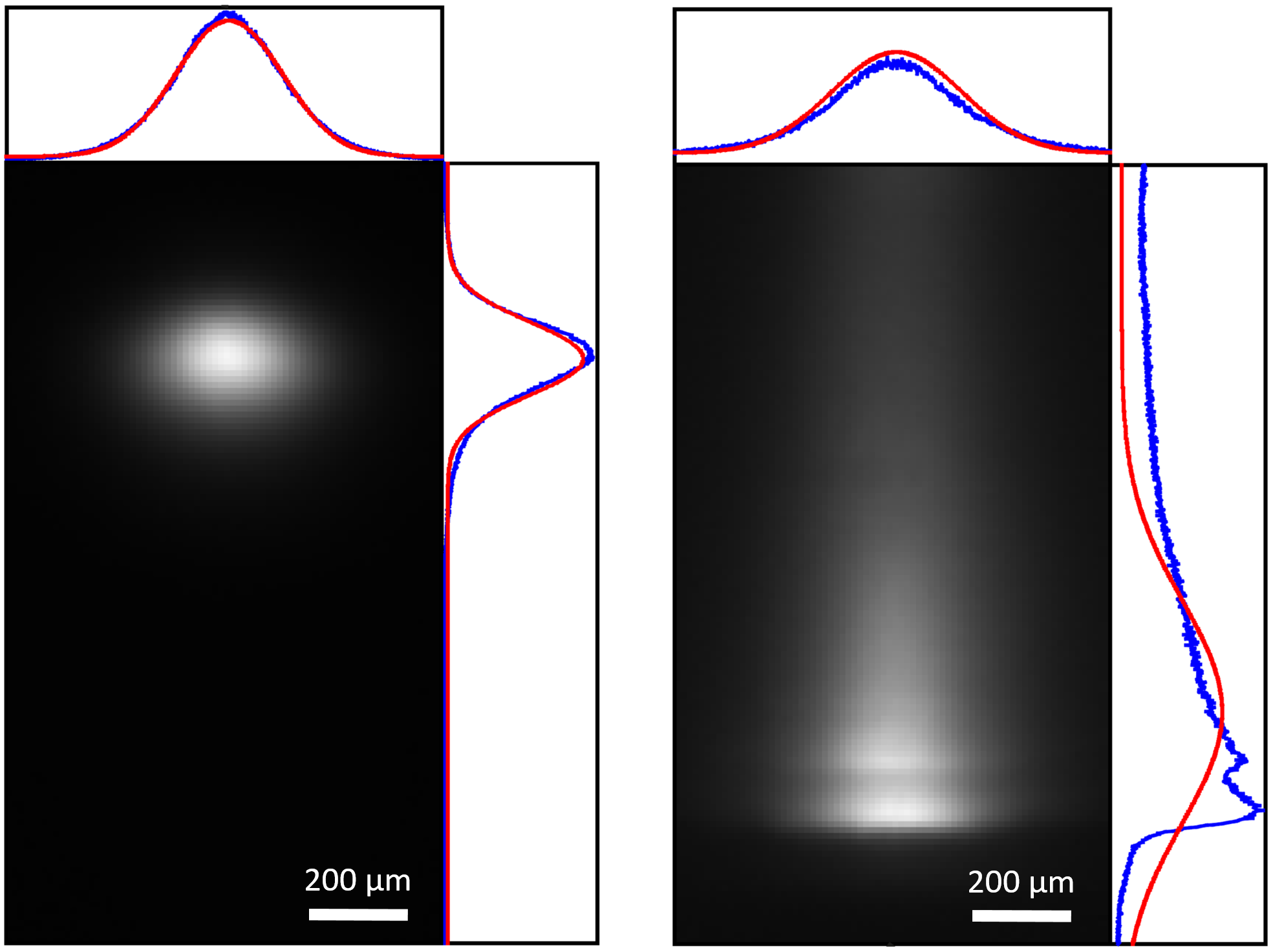}
  \caption{Time of flight images of the cloud (left) close to the optimum parameters, the cloud appears Gaussian, as expected from a thermal ensemble. (right) Far from the optimal conditions, caustics appear.}
\end{figure}  indicating that some gravitational energy was not compensated by the finite fall, and that the ensemble is not fully equilibrated at this stage. Indeed, the time of flight images reveal near-thermal distributions for the optimal trajectory and non-thermal distributions far from optimum, including the emergence of caustics which are characteristic of non-adiabatically falling clouds~\cite{Rosenblum2014}.\newline
From the geometric average $T=(T_x^{2}T_y)^{1/3}=64~\rm\mu K$, $N=1.5\times10^8$ and the trap gradient, the PSD is calculated to be $5.3\times10^{-5}$. This corresponds to about a factor of two smaller than the simulated result of $10\times10^{-5}$ (see table \ref{tab_psd}).
Some of the PSD loss is due to atom loss perhaps due to magnetic noises transferring atoms to the untrappable state. An additional possible explanation is the inaccuracy in the control of the currents. In order to find the PSD for a static trap we suddenly turn on the magnetic trap to a constant value which we also optimize.
In that case the PSD is $\sim 20$-fold smaller compared to the falling trap method but is still large compared to similar conventional setups, owing to the initial high PSD of our Raman sideband cooled atoms.
The thermodynamic parameters of the ensemble enable us to proceed with an efficient and fast microwave-forced evaporation, thus increasing the PSD by over a factor of 10 within 1 second. We are then limited by Majorana losses~\cite{Majorana1932} such that it is necessary to either "plug"~\cite{Davis1995} the quadrupole trap or to transfer the atoms into an optical dipole trap~\cite{Linn2009} in order to prevent losses and heating.



\emph{Conclusions - } We have shown that the use of a falling magnetic trap can compensate for large gravitational energies which would have otherwise led to heating. We experimentally demonstrated a 20-fold improvement of the PSD compared to a static magnetic trap method.
With the use of an RSBC stage and a falling trap we can start magnetic evaporation from a phase space density $\sim100$ times higher than in conventional apparatus, allowing to save time and reduce atom loss. For conventional Sisyphus cooling, the kinetic energy is comparable to the gravitational energy, so that phase-space density can be maintained when loading into a static magnetic trap. The concept of loading into a falling magnetic trap can be extended to other types of traps such as optical traps whose position and gradients can be easily controlled~\cite{Grimm2000}.

\emph{Acknowledgements} The authors would like to thank Yoav Sagi for his help. This work was supported in part by the ICore Israeli excellence center Circle of light and the Weizmann Institute Texas A\& M collaboration program.



\bibliographystyle{apsrev4-1}
\bibliography{bibliography}

\begin{thebibliography}{24}%
\makeatletter
\providecommand \@ifxundefined [1]{%
 \@ifx{#1\undefined}
}%
\providecommand \@ifnum [1]{%
 \ifnum #1\expandafter \@firstoftwo
 \else \expandafter \@secondoftwo
 \fi
}%
\providecommand \@ifx [1]{%
 \ifx #1\expandafter \@firstoftwo
 \else \expandafter \@secondoftwo
 \fi
}%
\providecommand \natexlab [1]{#1}%
\providecommand \enquote  [1]{``#1''}%
\providecommand \bibnamefont  [1]{#1}%
\providecommand \bibfnamefont [1]{#1}%
\providecommand \citenamefont [1]{#1}%
\providecommand \href@noop [0]{\@secondoftwo}%
\providecommand \href [0]{\begingroup \@sanitize@url \@href}%
\providecommand \@href[1]{\@@startlink{#1}\@@href}%
\providecommand \@@href[1]{\endgroup#1\@@endlink}%
\providecommand \@sanitize@url [0]{\catcode `\\12\catcode `\$12\catcode
  `\&12\catcode `\#12\catcode `\^12\catcode `\_12\catcode `\%12\relax}%
\providecommand \@@startlink[1]{}%
\providecommand \@@endlink[0]{}%
\providecommand \url  [0]{\begingroup\@sanitize@url \@url }%
\providecommand \@url [1]{\endgroup\@href {#1}{\urlprefix }}%
\providecommand \urlprefix  [0]{URL }%
\providecommand \Eprint [0]{\href }%
\providecommand \doibase [0]{http://dx.doi.org/}%
\providecommand \selectlanguage [0]{\@gobble}%
\providecommand \bibinfo  [0]{\@secondoftwo}%
\providecommand \bibfield  [0]{\@secondoftwo}%
\providecommand \translation [1]{[#1]}%
\providecommand \BibitemOpen [0]{}%
\providecommand \bibitemStop [0]{}%
\providecommand \bibitemNoStop [0]{.\EOS\space}%
\providecommand \EOS [0]{\spacefactor3000\relax}%
\providecommand \BibitemShut  [1]{\csname bibitem#1\endcsname}%
\let\auto@bib@innerbib\@empty
\bibitem [{\citenamefont {Anderson}\ \emph {et~al.}(1995)\citenamefont
  {Anderson}, \citenamefont {Ensher}, \citenamefont {Matthews}, \citenamefont
  {Wieman},\ and\ \citenamefont {Cornell}}]{Anderson1995}%
  \BibitemOpen
  \bibfield  {author} {\bibinfo {author} {\bibfnamefont {M.~H.}\ \bibnamefont
  {Anderson}}, \bibinfo {author} {\bibfnamefont {J.~R.}\ \bibnamefont
  {Ensher}}, \bibinfo {author} {\bibfnamefont {M.~R.}\ \bibnamefont
  {Matthews}}, \bibinfo {author} {\bibfnamefont {C.~E.}\ \bibnamefont
  {Wieman}}, \ and\ \bibinfo {author} {\bibfnamefont {E.~A.}\ \bibnamefont
  {Cornell}},\ }\href {\doibase 10.1126/science.269.5221.198} {\bibfield
  {journal} {\bibinfo  {journal} {Science}\ }\textbf {\bibinfo {volume}
  {269}},\ \bibinfo {pages} {198} (\bibinfo {year} {1995})}\BibitemShut
  {NoStop}%
\bibitem [{\citenamefont {Pethick}\ and\ \citenamefont
  {Smith}(2008)}]{pethick_smith_2008}%
  \BibitemOpen
  \bibfield  {author} {\bibinfo {author} {\bibfnamefont {C.~J.}\ \bibnamefont
  {Pethick}}\ and\ \bibinfo {author} {\bibfnamefont {H.}~\bibnamefont
  {Smith}},\ }\href {\doibase 10.1017/CBO9780511802850} {\emph {\bibinfo
  {title} {Bose Einstein Condensation in Dilute Gases}}},\ \bibinfo {edition}
  {2nd}\ ed.\ (\bibinfo  {publisher} {Cambridge University Press},\ \bibinfo
  {year} {2008})\BibitemShut {NoStop}%
\bibitem [{\citenamefont {Truscott}\ \emph {et~al.}(2001)\citenamefont
  {Truscott}, \citenamefont {Strecker}, \citenamefont {McAlexander},
  \citenamefont {Partridge},\ and\ \citenamefont {Hulet}}]{Truscott2001}%
  \BibitemOpen
  \bibfield  {author} {\bibinfo {author} {\bibfnamefont {A.~G.}\ \bibnamefont
  {Truscott}}, \bibinfo {author} {\bibfnamefont {K.~E.}\ \bibnamefont
  {Strecker}}, \bibinfo {author} {\bibfnamefont {W.~I.}\ \bibnamefont
  {McAlexander}}, \bibinfo {author} {\bibfnamefont {G.~B.}\ \bibnamefont
  {Partridge}}, \ and\ \bibinfo {author} {\bibfnamefont {R.~G.}\ \bibnamefont
  {Hulet}},\ }\href {\doibase 10.1126/science.1059318} {\bibfield  {journal}
  {\bibinfo  {journal} {Science}\ }\textbf {\bibinfo {volume} {291}},\ \bibinfo
  {pages} {2570} (\bibinfo {year} {2001})}\BibitemShut {NoStop}%
\bibitem [{\citenamefont {Bloch}(2005)}]{Bloch2005}%
  \BibitemOpen
  \bibfield  {author} {\bibinfo {author} {\bibfnamefont {I.}~\bibnamefont
  {Bloch}},\ }\href {\doibase 10.1038/nphys138} {\bibfield  {journal} {\bibinfo
   {journal} {Nat Phys}\ }\textbf {\bibinfo {volume} {1}},\ \bibinfo {pages}
  {23} (\bibinfo {year} {2005})}\BibitemShut {NoStop}%
\bibitem [{\citenamefont {Bloch}\ \emph {et~al.}(2012)\citenamefont {Bloch},
  \citenamefont {Dalibard},\ and\ \citenamefont {Nascimbene}}]{Bloch2012}%
  \BibitemOpen
  \bibfield  {author} {\bibinfo {author} {\bibfnamefont {I.}~\bibnamefont
  {Bloch}}, \bibinfo {author} {\bibfnamefont {J.}~\bibnamefont {Dalibard}}, \
  and\ \bibinfo {author} {\bibfnamefont {S.}~\bibnamefont {Nascimbene}},\
  }\href {\doibase 10.1038/nphys2259} {\bibfield  {journal} {\bibinfo
  {journal} {Nat Phys}\ }\textbf {\bibinfo {volume} {8}},\ \bibinfo {pages}
  {267} (\bibinfo {year} {2012})}\BibitemShut {NoStop}%
\bibitem [{\citenamefont {Greiner}\ \emph {et~al.}(2002)\citenamefont
  {Greiner}, \citenamefont {Mandel}, \citenamefont {Esslinger}, \citenamefont
  {Hansch},\ and\ \citenamefont {Bloch}}]{Greiner2002}%
  \BibitemOpen
  \bibfield  {author} {\bibinfo {author} {\bibfnamefont {M.}~\bibnamefont
  {Greiner}}, \bibinfo {author} {\bibfnamefont {O.}~\bibnamefont {Mandel}},
  \bibinfo {author} {\bibfnamefont {T.}~\bibnamefont {Esslinger}}, \bibinfo
  {author} {\bibfnamefont {T.~W.}\ \bibnamefont {Hansch}}, \ and\ \bibinfo
  {author} {\bibfnamefont {I.}~\bibnamefont {Bloch}},\ }\href {\doibase
  10.1038/415039a} {\bibfield  {journal} {\bibinfo  {journal} {Nature}\
  }\textbf {\bibinfo {volume} {415}},\ \bibinfo {pages} {39} (\bibinfo {year}
  {2002})}\BibitemShut {NoStop}%
\bibitem [{\citenamefont {Goldman}\ \emph {et~al.}(2016)\citenamefont
  {Goldman}, \citenamefont {Budich},\ and\ \citenamefont
  {Zoller}}]{Goldman2016}%
  \BibitemOpen
  \bibfield  {author} {\bibinfo {author} {\bibfnamefont {N.}~\bibnamefont
  {Goldman}}, \bibinfo {author} {\bibfnamefont {J.~C.}\ \bibnamefont {Budich}},
  \ and\ \bibinfo {author} {\bibfnamefont {P.}~\bibnamefont {Zoller}},\ }\href
  {http://dx.doi.org/10.1038/nphys3803} {\bibfield  {journal} {\bibinfo
  {journal} {Nat Phys}\ }\textbf {\bibinfo {volume} {12}},\ \bibinfo {pages}
  {639} (\bibinfo {year} {2016})},\ \bibinfo {note} {progress
  Article}\BibitemShut {NoStop}%
\bibitem [{\citenamefont {Phillips}(1997)}]{Phillips1997}%
  \BibitemOpen
  \bibfield  {author} {\bibinfo {author} {\bibfnamefont {W.}~\bibnamefont
  {Phillips}},\ }\href@noop {} {\bibfield  {journal} {\bibinfo  {journal}
  {Nobel Lectures}\ } (\bibinfo {year} {1997})}\BibitemShut {NoStop}%
\bibitem [{\citenamefont {Kerman}\ \emph {et~al.}(2000)\citenamefont {Kerman},
  \citenamefont {Vuleti\ifmmode~\acute{c}\else \'{c}\fi{}}, \citenamefont
  {Chin},\ and\ \citenamefont {Chu}}]{Kerman2000}%
  \BibitemOpen
  \bibfield  {author} {\bibinfo {author} {\bibfnamefont {A.~J.}\ \bibnamefont
  {Kerman}}, \bibinfo {author} {\bibfnamefont {V.}~\bibnamefont
  {Vuleti\ifmmode~\acute{c}\else \'{c}\fi{}}}, \bibinfo {author} {\bibfnamefont
  {C.}~\bibnamefont {Chin}}, \ and\ \bibinfo {author} {\bibfnamefont
  {S.}~\bibnamefont {Chu}},\ }\href {\doibase 10.1103/PhysRevLett.84.439}
  {\bibfield  {journal} {\bibinfo  {journal} {Phys. Rev. Lett.}\ }\textbf
  {\bibinfo {volume} {84}},\ \bibinfo {pages} {439} (\bibinfo {year}
  {2000})}\BibitemShut {NoStop}%
\bibitem [{\citenamefont {Migdall}\ \emph {et~al.}(1985)\citenamefont
  {Migdall}, \citenamefont {Prodan}, \citenamefont {Phillips}, \citenamefont
  {Bergeman},\ and\ \citenamefont {Metcalf}}]{Migdall1985}%
  \BibitemOpen
  \bibfield  {author} {\bibinfo {author} {\bibfnamefont {A.~L.}\ \bibnamefont
  {Migdall}}, \bibinfo {author} {\bibfnamefont {J.~V.}\ \bibnamefont {Prodan}},
  \bibinfo {author} {\bibfnamefont {W.~D.}\ \bibnamefont {Phillips}}, \bibinfo
  {author} {\bibfnamefont {T.~H.}\ \bibnamefont {Bergeman}}, \ and\ \bibinfo
  {author} {\bibfnamefont {H.~J.}\ \bibnamefont {Metcalf}},\ }\href {\doibase
  10.1103/PhysRevLett.54.2596} {\bibfield  {journal} {\bibinfo  {journal}
  {Phys. Rev. Lett.}\ }\textbf {\bibinfo {volume} {54}},\ \bibinfo {pages}
  {2596} (\bibinfo {year} {1985})}\BibitemShut {NoStop}%
\bibitem [{\citenamefont {Ketterle}\ \emph {et~al.}(1999)\citenamefont
  {Ketterle}, \citenamefont {Durfee},\ and\ \citenamefont
  {Stamper-kurn}}]{Ketterle1999}%
  \BibitemOpen
  \bibfield  {author} {\bibinfo {author} {\bibfnamefont {W.}~\bibnamefont
  {Ketterle}}, \bibinfo {author} {\bibfnamefont {D.~S.}\ \bibnamefont
  {Durfee}}, \ and\ \bibinfo {author} {\bibfnamefont {D.~M.}\ \bibnamefont
  {Stamper-kurn}},\ }in\ \href@noop {} {\emph {\bibinfo {booktitle}
  {Proceedings of the international school of physics ”Enrico Fermi”}}}\
  (\bibinfo  {publisher} {Press},\ \bibinfo {year} {1999})\BibitemShut
  {NoStop}%
\bibitem [{\citenamefont {Kasevich}\ and\ \citenamefont
  {Chu}(1992)}]{Kasevich1992}%
  \BibitemOpen
  \bibfield  {author} {\bibinfo {author} {\bibfnamefont {M.}~\bibnamefont
  {Kasevich}}\ and\ \bibinfo {author} {\bibfnamefont {S.}~\bibnamefont {Chu}},\
  }\href {\doibase 10.1103/PhysRevLett.69.1741} {\bibfield  {journal} {\bibinfo
   {journal} {Phys. Rev. Lett.}\ }\textbf {\bibinfo {volume} {69}},\ \bibinfo
  {pages} {1741} (\bibinfo {year} {1992})}\BibitemShut {NoStop}%
\bibitem [{\citenamefont {Davidson}\ \emph {et~al.}(1994)\citenamefont
  {Davidson}, \citenamefont {Lee}, \citenamefont {Kasevich},\ and\
  \citenamefont {Chu}}]{Davidson1994}%
  \BibitemOpen
  \bibfield  {author} {\bibinfo {author} {\bibfnamefont {N.}~\bibnamefont
  {Davidson}}, \bibinfo {author} {\bibfnamefont {H.~J.}\ \bibnamefont {Lee}},
  \bibinfo {author} {\bibfnamefont {M.}~\bibnamefont {Kasevich}}, \ and\
  \bibinfo {author} {\bibfnamefont {S.}~\bibnamefont {Chu}},\ }\href {\doibase
  10.1103/PhysRevLett.72.3158} {\bibfield  {journal} {\bibinfo  {journal}
  {Phys. Rev. Lett.}\ }\textbf {\bibinfo {volume} {72}},\ \bibinfo {pages}
  {3158} (\bibinfo {year} {1994})}\BibitemShut {NoStop}%
\bibitem [{\citenamefont {Aspect}\ \emph {et~al.}(1988)\citenamefont {Aspect},
  \citenamefont {Arimondo}, \citenamefont {Kaiser}, \citenamefont
  {Vansteenkiste},\ and\ \citenamefont {Cohen-Tannoudji}}]{Aspect1988}%
  \BibitemOpen
  \bibfield  {author} {\bibinfo {author} {\bibfnamefont {A.}~\bibnamefont
  {Aspect}}, \bibinfo {author} {\bibfnamefont {E.}~\bibnamefont {Arimondo}},
  \bibinfo {author} {\bibfnamefont {R.}~\bibnamefont {Kaiser}}, \bibinfo
  {author} {\bibfnamefont {N.}~\bibnamefont {Vansteenkiste}}, \ and\ \bibinfo
  {author} {\bibfnamefont {C.}~\bibnamefont {Cohen-Tannoudji}},\ }\href
  {\doibase 10.1103/PhysRevLett.61.826} {\bibfield  {journal} {\bibinfo
  {journal} {Phys. Rev. Lett.}\ }\textbf {\bibinfo {volume} {61}},\ \bibinfo
  {pages} {826} (\bibinfo {year} {1988})}\BibitemShut {NoStop}%
\bibitem [{\citenamefont {Aspect}\ \emph {et~al.}(1989)\citenamefont {Aspect},
  \citenamefont {Arimondo}, \citenamefont {Kaiser}, \citenamefont
  {Vansteenkiste},\ and\ \citenamefont {Cohen-Tannoudji}}]{Aspect1989}%
  \BibitemOpen
  \bibfield  {author} {\bibinfo {author} {\bibfnamefont {A.}~\bibnamefont
  {Aspect}}, \bibinfo {author} {\bibfnamefont {E.}~\bibnamefont {Arimondo}},
  \bibinfo {author} {\bibfnamefont {R.}~\bibnamefont {Kaiser}}, \bibinfo
  {author} {\bibfnamefont {N.}~\bibnamefont {Vansteenkiste}}, \ and\ \bibinfo
  {author} {\bibfnamefont {C.}~\bibnamefont {Cohen-Tannoudji}},\ }\href
  {\doibase 10.1364/JOSAB.6.002112} {\bibfield  {journal} {\bibinfo  {journal}
  {J. Opt. Soc. Am. B}\ }\textbf {\bibinfo {volume} {6}},\ \bibinfo {pages}
  {2112} (\bibinfo {year} {1989})}\BibitemShut {NoStop}%
\bibitem [{\citenamefont {Streed}\ \emph {et~al.}(2006)\citenamefont {Streed},
  \citenamefont {Chikkatur}, \citenamefont {Gustavson}, \citenamefont {Boyd},
  \citenamefont {Torii}, \citenamefont {Schneble}, \citenamefont {Campbell},
  \citenamefont {Pritchard},\ and\ \citenamefont {Ketterle}}]{streed2006}%
  \BibitemOpen
  \bibfield  {author} {\bibinfo {author} {\bibfnamefont {E.~W.}\ \bibnamefont
  {Streed}}, \bibinfo {author} {\bibfnamefont {A.~P.}\ \bibnamefont
  {Chikkatur}}, \bibinfo {author} {\bibfnamefont {T.~L.}\ \bibnamefont
  {Gustavson}}, \bibinfo {author} {\bibfnamefont {M.}~\bibnamefont {Boyd}},
  \bibinfo {author} {\bibfnamefont {Y.}~\bibnamefont {Torii}}, \bibinfo
  {author} {\bibfnamefont {D.}~\bibnamefont {Schneble}}, \bibinfo {author}
  {\bibfnamefont {G.~K.}\ \bibnamefont {Campbell}}, \bibinfo {author}
  {\bibfnamefont {D.~E.}\ \bibnamefont {Pritchard}}, \ and\ \bibinfo {author}
  {\bibfnamefont {W.}~\bibnamefont {Ketterle}},\ }\href {\doibase
  10.1063/1.2163977} {\bibfield  {journal} {\bibinfo  {journal} {Review of
  Scientific Instruments}\ }\textbf {\bibinfo {volume} {77}},\ \bibinfo {pages}
  {023106} (\bibinfo {year} {2006})}\BibitemShut {NoStop}%
\bibitem [{\citenamefont {Lin}\ \emph {et~al.}(2009)\citenamefont {Lin},
  \citenamefont {Perry}, \citenamefont {Compton}, \citenamefont {Spielman},\
  and\ \citenamefont {Porto}}]{Linn2009}%
  \BibitemOpen
  \bibfield  {author} {\bibinfo {author} {\bibfnamefont {Y.-J.}\ \bibnamefont
  {Lin}}, \bibinfo {author} {\bibfnamefont {A.~R.}\ \bibnamefont {Perry}},
  \bibinfo {author} {\bibfnamefont {R.~L.}\ \bibnamefont {Compton}}, \bibinfo
  {author} {\bibfnamefont {I.~B.}\ \bibnamefont {Spielman}}, \ and\ \bibinfo
  {author} {\bibfnamefont {J.~V.}\ \bibnamefont {Porto}},\ }\href {\doibase
  10.1103/PhysRevA.79.063631} {\bibfield  {journal} {\bibinfo  {journal} {Phys.
  Rev. A}\ }\textbf {\bibinfo {volume} {79}},\ \bibinfo {pages} {063631}
  (\bibinfo {year} {2009})}\BibitemShut {NoStop}%
\bibitem [{\citenamefont {Lavert-Ofir}\ \emph {et~al.}(2011)\citenamefont
  {Lavert-Ofir}, \citenamefont {Gersten}, \citenamefont {Henson}, \citenamefont
  {Shani}, \citenamefont {David}, \citenamefont {Narevicius},\ and\
  \citenamefont {Narevicius}}]{Edvardas2011}%
  \BibitemOpen
  \bibfield  {author} {\bibinfo {author} {\bibfnamefont {E.}~\bibnamefont
  {Lavert-Ofir}}, \bibinfo {author} {\bibfnamefont {S.}~\bibnamefont
  {Gersten}}, \bibinfo {author} {\bibfnamefont {A.~B.}\ \bibnamefont {Henson}},
  \bibinfo {author} {\bibfnamefont {I.}~\bibnamefont {Shani}}, \bibinfo
  {author} {\bibfnamefont {L.}~\bibnamefont {David}}, \bibinfo {author}
  {\bibfnamefont {J.}~\bibnamefont {Narevicius}}, \ and\ \bibinfo {author}
  {\bibfnamefont {E.}~\bibnamefont {Narevicius}},\ }\href
  {http://stacks.iop.org/1367-2630/13/i=10/a=103030} {\bibfield  {journal}
  {\bibinfo  {journal} {New Journal of Physics}\ }\textbf {\bibinfo {volume}
  {13}},\ \bibinfo {pages} {103030} (\bibinfo {year} {2011})}\BibitemShut
  {NoStop}%
\bibitem [{\citenamefont {Sagi}(2012)}]{Sagi2012}%
  \BibitemOpen
  \bibfield  {author} {\bibinfo {author} {\bibfnamefont {Y.}~\bibnamefont
  {Sagi}},\ }\href {\doibase 10.1007/978-3-642-29605-5} {\emph {\bibinfo
  {title} {Collisional Narrowing and Dynamical Decoupling in a Dense Ensemble
  of Cold Atoms}}},\ \bibinfo {edition} {1st}\ ed.\ (\bibinfo  {publisher}
  {Springer-Verlag Berlin Heidelberg},\ \bibinfo {year} {2012})\BibitemShut
  {NoStop}%
\bibitem [{\citenamefont {Dalibard}\ and\ \citenamefont
  {Cohen-Tannoudji}(1989)}]{Dalibard1989}%
  \BibitemOpen
  \bibfield  {author} {\bibinfo {author} {\bibfnamefont {J.}~\bibnamefont
  {Dalibard}}\ and\ \bibinfo {author} {\bibfnamefont {C.}~\bibnamefont
  {Cohen-Tannoudji}},\ }\href {\doibase 10.1364/JOSAB.6.002023} {\bibfield
  {journal} {\bibinfo  {journal} {Journal of the Optical Society of America B}\
  }\textbf {\bibinfo {volume} {6}},\ \bibinfo {pages} {2023} (\bibinfo {year}
  {1989})}\BibitemShut {NoStop}%
\bibitem [{\citenamefont {Rosenblum}\ \emph {et~al.}(2014)\citenamefont
  {Rosenblum}, \citenamefont {Bechler}, \citenamefont {Shomroni}, \citenamefont
  {Kaner}, \citenamefont {Arusi-Parpar}, \citenamefont {Raz},\ and\
  \citenamefont {Dayan}}]{Rosenblum2014}%
  \BibitemOpen
  \bibfield  {author} {\bibinfo {author} {\bibfnamefont {S.}~\bibnamefont
  {Rosenblum}}, \bibinfo {author} {\bibfnamefont {O.}~\bibnamefont {Bechler}},
  \bibinfo {author} {\bibfnamefont {I.}~\bibnamefont {Shomroni}}, \bibinfo
  {author} {\bibfnamefont {R.}~\bibnamefont {Kaner}}, \bibinfo {author}
  {\bibfnamefont {T.}~\bibnamefont {Arusi-Parpar}}, \bibinfo {author}
  {\bibfnamefont {O.}~\bibnamefont {Raz}}, \ and\ \bibinfo {author}
  {\bibfnamefont {B.}~\bibnamefont {Dayan}},\ }\href {\doibase
  10.1103/PhysRevLett.112.120403} {\bibfield  {journal} {\bibinfo  {journal}
  {Phys. Rev. Lett.}\ }\textbf {\bibinfo {volume} {112}},\ \bibinfo {pages}
  {120403} (\bibinfo {year} {2014})}\BibitemShut {NoStop}%
\bibitem [{\citenamefont {Majorana}(1932)}]{Majorana1932}%
  \BibitemOpen
  \bibfield  {author} {\bibinfo {author} {\bibfnamefont {E.}~\bibnamefont
  {Majorana}},\ }\href {\doibase 10.1007/BF02960953} {\bibfield  {journal}
  {\bibinfo  {journal} {Il Nuovo Cimento (1924-1942)}\ }\textbf {\bibinfo
  {volume} {9}},\ \bibinfo {pages} {43} (\bibinfo {year} {1932})}\BibitemShut
  {NoStop}%
\bibitem [{\citenamefont {Davis}\ \emph {et~al.}(1995)\citenamefont {Davis},
  \citenamefont {Mewes}, \citenamefont {Andrews}, \citenamefont {van Druten},
  \citenamefont {Durfee}, \citenamefont {Kurn},\ and\ \citenamefont
  {Ketterle}}]{Davis1995}%
  \BibitemOpen
  \bibfield  {author} {\bibinfo {author} {\bibfnamefont {K.~B.}\ \bibnamefont
  {Davis}}, \bibinfo {author} {\bibfnamefont {M.~O.}\ \bibnamefont {Mewes}},
  \bibinfo {author} {\bibfnamefont {M.~R.}\ \bibnamefont {Andrews}}, \bibinfo
  {author} {\bibfnamefont {N.~J.}\ \bibnamefont {van Druten}}, \bibinfo
  {author} {\bibfnamefont {D.~S.}\ \bibnamefont {Durfee}}, \bibinfo {author}
  {\bibfnamefont {D.~M.}\ \bibnamefont {Kurn}}, \ and\ \bibinfo {author}
  {\bibfnamefont {W.}~\bibnamefont {Ketterle}},\ }\href {\doibase
  10.1103/PhysRevLett.75.3969} {\bibfield  {journal} {\bibinfo  {journal}
  {Phys. Rev. Lett.}\ }\textbf {\bibinfo {volume} {75}},\ \bibinfo {pages}
  {3969} (\bibinfo {year} {1995})}\BibitemShut {NoStop}%
\bibitem [{\citenamefont {Grimm}\ \emph {et~al.}(2000)\citenamefont {Grimm},
  \citenamefont {Weidemüller},\ and\ \citenamefont {Ovchinnikov}}]{Grimm2000}%
  \BibitemOpen
  \bibfield  {author} {\bibinfo {author} {\bibfnamefont {R.}~\bibnamefont
  {Grimm}}, \bibinfo {author} {\bibfnamefont {M.}~\bibnamefont {Weidemüller}},
  \ and\ \bibinfo {author} {\bibfnamefont {Y.~B.}\ \bibnamefont
  {Ovchinnikov}},\ }\href {\doibase
  http://dx.doi.org/10.1016/S1049-250X(08)60186-X} {\bibfield  {journal}
  {\bibinfo  {journal} {Advances In Atomic, Molecular, and Optical Physics}\
  }\textbf {\bibinfo {volume} {42}},\ \bibinfo {pages} {95 } (\bibinfo {year}
  {2000})}\BibitemShut {NoStop}%
\end{thebibliography}%



\end{document}